# Dark photon search with a gyrotron and a transition edge sensor

A. Miyazaki[1], P. Spagnolo[2]
[1]Uppsala University, Uppsala, Sweden
[2]Istituto Nazionale Fisica Nucleare, Pisa, Italy

*Abstract*—A dark photon, one of the candidates of light dark matter, will be searched around 0.1 meV range by using a gyrotron. The use of a Transition Edge Sensor is the key of this experiment and the expected result is promising. This search will pave a way to future axion search using similar instruments.

## I. INTRODUCTION

Dark matter is a missing piece of modern particle physics while its existence is strongly suggested by astrophysics and cosmology. Recent collider and direct search experiments have been excluding the heavy dark matter candidates with a mass range of above eV [1]. Thus, the search for light dark matter candidates below eV will be more and more motivated in the post Large Hadron Collider (LHC) era.

One promising candidate of light dark matter is a massive extra U(1) gauge boson, called dark photon [2]. This new particle beyond the Standard Model can mix to one ordinary photon and its existence could be addressed with a high-power photon source and a sensitive photon detector. There have been various experiments, using a Light-Shining-Through-the-Wall (LSW) technique, to find dark photons with optical laser [3], X-rays [4], Radio Frequency of below 10 GHz [5]. No clear indication of the dark photons has been observed with these frequency regions. A complementary experiment with a frequency that has never been addressed is therefore motivated. Especially, the well-known THz-gap [6] provides a promising opportunity on this subject with reasonable technical challenges, which can be resolved within a few years.

In 2012, the author proposed a new experiment for dark photons [7] using a gyrotron at 200 GHz, developed for the spectroscopy of hyperfine structure of positronium [8, 9, 10]. We proposed the use of superconducting heterodyne mixer (SIS-mixer) [11] operated at 4.2 K, originally developed to detect astrophysical millimeter waves in the ALMA observatory [12]. The Noise Equivalent Power (NEP) of that sensor was evaluated to be $6\times10^{-14}$ W/√MHz, from which we could drastically reduce noise levels by selecting the narrow-band region around the gyrotron signal. However, this experiment was not successfully implemented because our gyrotron [13] had a spectral band-width of between 1 MHz and 10 MHz when the power was in the kW level. This was exceptionally narrow in the millimeter-wave range but was still not sufficient to obtain the desired noise levels for dark photons search. Another issue was that the SIS-mixer was optimized for higher frequency above 200 GHz but lower frequency below 200 GHz was preferred for the dark photon search to reduce the intrinsic background noise from the black body radiation. This sensor reached a quantum noise limit already at 4.5 K and no further improvement was feasible at that moment.

In this new paper, the author proposes the use of a Transition Edge Sensor (TES) operated inside a dilution refrigerator at 20 mK. As shown in Fig. 1, National Enterprise for nano Science and Nano Technology (NEST) and INFN Pisa, in Italy, developed a TES above 30 GHz dedicated to the STAX (Axion-like particle searches with sub-THz photons) collaboration [14]. The expected NEP is $10^{-17}$ W/√Hz with blackbody radiation from 300 K, and $10^{-20}$ W/√Hz with a thermally shielded environment. The sensitivity clearly exceeds the previous proposal of mine, with such a dedicated instrument for millimeter-wave particle physics. In this experiment, we integrate two promising scientific instruments, a high-power gyrotron and a TES, to make the first search for the dark photons in the millimeter-wave range.

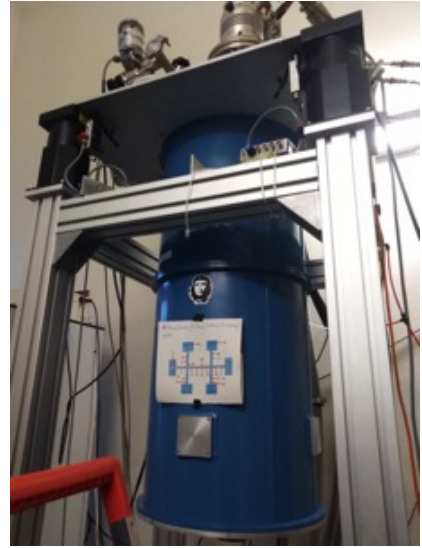

**Fig. 1.** Dilution refrigerator for the TES dedicated to the dark photons and ALPs search in NEST, Pisa, Italy

## II. EXPECTED SEARCH RANGE

The exclusion limit was estimated as shown in Fig. 2. We estimate the measurement sensitivity out of the state-of-the-art gyrotrons in Karlsruher Institut für Technologie (KIT). The black line shows a conservative scenario with a TES of $10^{-17}$ W/√Hz and a CW gyrotron at 17 kW and 28 GHz. The operation time was assumed to be half an hour. This can already address the gap between previous experiments by RF and optical laser. The red curve shows an optimistic case with a TES with a 100 mK wall in front to reach $10^{-20}$ W/√Hz and a CW gyrotron at 17 kW and 140 GHz. A MW class gyrotron will even improve the sensitivity by a factor of three. The experiment will be carried out within three years.

Electrical and radiation noise could be additional background noises to this experiment. Just as addressed in the previous experiments [8], appropriate shielding and grounding of all the cables will be sufficient. In addition, the sensitive photon sensor and electronics will be installed in the dilution refrigerator (Fig.1), which will act as an ideal Faraday cage when it is completely closed.

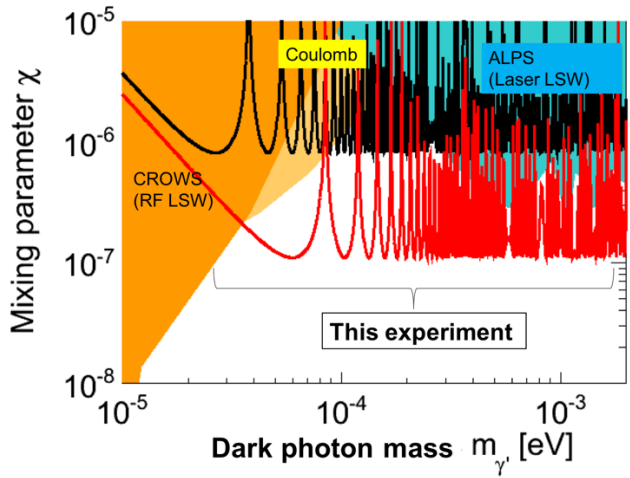

**Fig. 2.** Expected search range of the dark photons with a conservative case (black curve) and optimistic case (red) together with other laboratory-based experiments [3, 5, 15]. The Coulomb and ALPS plots are courtesy of T. Inada (The University of Tokyo).

### III. TOWARD THE AXION SEARCH

Another important yet-to-discover particle beyond the Standard Model of particle physics is the axion, originally introduced as a byproduct of a natural solution of the strong CP problem in the Quantum Chromodynamics [16]. Recent advances in the super string theory naturally predict [17] a bunch of particles which have a similar property of such a classic axion. These particles are in general called Axion-Like Particles (APLs). Although the dark photon mixes to one ordinary photon, APLs couple with two photons. Typical laboratory-based searches for APLs are therefore composed of a photon device with a strong magnetic field as a second off-shell photon source. In principle, the setup to be developed for the dark photon search will be compatible with the ALPs search once dipole magnets are integrated to the experiment.

In future experiment by the STAX collaboration, the used of a prototype $Nb_3Sn$ dipole magnet of 11 T developed for the High-Luminosity LHC is currently under consideration. The bore diameter of this magnet is a CERN standard 60 mm and the length is 1.5 m. The $TEM_{00}$ mode in a free space (Gaussian beam) developed in the previous experiments [9] cannot be confined in such a narrow space because of the short Rayleigh length. In addition, a free space mode might dissipate some power at the magnet and could trigger a magnet quench.

We propose a novel idea to use the $HE_{11}$ mode in a corrugated waveguide structure. The cavities of $HE_{11}$ mode have been considered for the microwave undulators [18, 19] at 12 GHz and 36 GHz. Their dimensions and experimental results would fulfill our requirement. Figure 3 shows the preliminary simulation (by CST[20]) results of a $HE_{11}$ cavity for the ALPs search. The linearly polarized electric fields mimic the conventional experiments in the Fabry-Pérot cavities in the laser experiment for ALPs. The corrugation also avoids the substantial heat load at the cavity wall.

Further optimization for the geometrical coupling factor and prototyping will be performed in coming years. One technical challenge is the thermal simulation and experiments to assess the maximum averaged power and duty ratio that this cavity can store as its internal energy, which is proportional to the number of photons and thus that of the hypothetical ALPs.

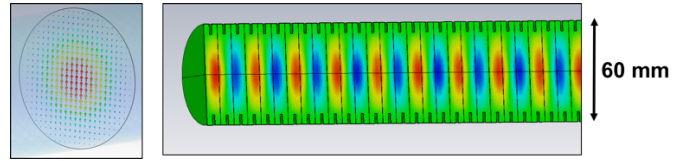

**Fig. 3.** $HE_{11}$ mode in a corrugated waveguide cavity for ALPs production and detection to be installed in an 11T $Nb_3Sn$ dipole magnet for HL-LHC.

### IV. CONCLUSION

The dark photon search is proposed with a gyrotron and a superconducting sensor. The major advance from previous experiment is the use of the TES in a dilution refrigerator. Their successful implementation will lead to the most sensitive laboratory-based experiment for the dark photon of mass range around 0.1 meV. We will also integrate the instruments with a $HE_{11}$ cavity inside a strong dipole magnet for the ALPs search. This experiment provides a complementary method as other experiments using lasers, RF, and X-rays. The whole project is a crucial milestone to establish the future particle physics by high-power millimeter waves in a laboratory.

### ACKNOWLEDGEMENTS

We gratefully acknowledge Dr. F. Giazotto for useful comments on the TES detector. Our sincere gratitude goes to Prof. M. Thumm for the discussion on the gyrotrons. The idea on the microwave cavity was supported by Prof. A. Cross. We also thank Ms. K. Narita, Dr. T. Inada, Prof. T. Namba, and Prof. S. Asai for their preliminary calculation of the dark photon search. Special thanks go to Dr. F. Caspers about discussion on this proposal in general.

### REFERENCES

[1] M. Tanabashi, et al. (Particle Data Group), Phys. Rev. D 98, 030001 (2018).
[2] P. Arias, et al, JCAP 1206 013 (2012).
[3] R Bähre, et al. (ALPS collaboration), JINST 8 T09001 (2013).
[4] T. Inada, et al., PLB, 722, 301-304 (2013).
[5] M. Betz, et al. (CROWS collaboration) Phys. Rev. D88:075014 (2013).
[6] G. P. Williams, Rep. Prog. Phys. 69 pp301-326 (2016).
[7] T. Suehara, K. Owada, A. Miyazaki, et al, DOI:10.1109/IRMMW-THz.2012.6380414, (2012).
[8] T. Yamazaki, A. Miyazaki, et al, Phys Rev Lett 108, 25, 253401 (2012).
[9] A. Miyazaki, et al, J. Infrared, Millimeter, and Terahertz Waves, 35, 1, pp.91-100, (2014).
[10] A. Miyazaki, et al, PTEP, 1, 011C01 (2015).
[11] V. P. Koshelets, et al., Appl. Phys. Lett. 68, 1273 (1996).
[12] https://www.almaobservatory.org/en/home/
[13] Y. Tatematsu et al., J. Infrared, Millimeter, and Terahertz Waves, 33, 3, pp.292-305 (2012).
[14] L. M. Capparelli, et al. Phys. Dark Univ. 12, 37 (2016).
[15] D. F. Bartlett and S. Lögl, Phys. Rev. Lett. 61, 2285 (1988).
[16] R. D. Peccei and H. R. Quinn, Phys. Rev. Lett. 38, 1440 (1977).
[17] A. Arvanitaki, et al., Phys. Rev. D 81, 123530 (2010).
[18] S. Tantawi et al, PRL112 164802 (2014).
[19] L. Zhang et al J. Synchrotron Rad. 26. 11-17 (2019).
[20] CST Microwave Studio, https://www.cst.com/products/cstmws